\documentclass{PoS}

\title{Monte Carlo Studies of Combined MAGIC and LST1 Observations}

\ShortTitle{MAGIC-LST1 combined observations}

\author{\speaker{F. Di Pierro}$^{1}$, L. Arrabito$^{2}$, A. Baquero Larriva$^{3}$, A. Berti$^{1}$, J. Bregeon$^{2}$, D. Depaoli$^{1,4}$, D. Dominis Prester$^{5}$, R. Lopez Coto$^{6}$, M. Manganaro$^{5}$, S. Mi\'canovi\'c$^{5}$, A. Moralejo$^{7}$, Y. Ohtani $^{8}$, L. Saha$^{3}$, J. Sitarek$^{9}$, Y. Suda$^{10}$, T. Terzi\'c$^{5}$, I. Vovk$^{10}$, T. Vuillaume$^{11}$ for the CTA Consortium\footnote{for consortium list see PoS(ICRC2019)1177}\\
        $^{1}$INFN Sezione di Torino, via P. Giuria, 1 - 10125 Torino, Italy\\
		  $^{2}$Laboratoire Univers et Particules de Montpellier, Universit\'e de Montpellier, CNRS/IN2P3, France\\
		  $^{3}$Universidad Complutense de Madrid, Avda Complutense s/n E-28040 Madrid, Spain\\
		  $^{4}$Universit\`a degli Studi di Torino, Dipartimento di Fisica, via P. Giuria, 1 - 10125 Torino, Italy\\
		  $^{5}$University of Rijeka, Department of Physics, Ul. Radmile Matejcic 2, HR-51000 Rijeka, Croatia\\
		  $^{6}$INFN Sezione di Padova, Via Marzolo, 8 - 35131 Padova, Italy\\
		  $^{7}$Institut de Fisica d'Altes Energies (IFAE), The Barcelona Institute of Science and Technology, Campus UAB, 08193 Bellaterra (Barcelona), Spain\\
		  $^{8}$ICRR, The University of Tokyo, Japan\\ 
		  $^{9}$Department of Astrophysics, The University of \L\'od\'z, ul. Pomorska 149/153, 90-236 \L\'od\'z, Poland\\
		  $^{10}$Max Planck Institute for Physics, München, Germany\\
		  $^{11}$LAPP, CNRS, Université Savoie Mont-Blanc, France\\
        E-mail: \email{federico.dipierro@to.infn.it}}

\abstract{The Cherenkov Telescope Array (CTA) is the next generation very high energy gamma-ray observatory covering the 20 GeV - 300 TeV energy range with unprecedented sensitivity, angular and energy resolution. With a site in each hemisphere, CTA will provide full-sky coverage. Four Large Size Telescopes (LSTs) in each site will be dedicated to the lowest energy range (20 GeV - 200 GeV). The first LST prototype has been installed at the CTA Northern site (Canary Island of La Palma, Spain) in October 2018 and it had been since then in commissioning phase. LST1 is located at about 100 m from MAGIC, a system of two 17m-diameter Imaging Atmospheric Cherenkov Telescopes designed to perform gamma-ray astronomy in the energy range from 50 GeV with standard trigger (30 GeV with SumTrigger) to 50 TeV and whose performance is very well established. The co-location of LST1 and MAGIC offers the great opportunity of cross-calibrating the two systems on an event-by-event basis. It will be indeed possible to compare the parameters of the same extensive air shower reconstructed by the two instruments. We investigated the performance that could be reached with combined observations.}

\FullConference{36th International Cosmic Ray Conference -ICRC2019-\\
		July 24th - August 1st, 2019\\
		Madison, WI, U.S.A.}

\begin{document}

\section{Introduction}

The Cherenkov Telescope Array (CTA\footnote{www.cta-observatory.org}) will observe very high energy gamma-rays with unprecedented performance \cite{gernot} from two sites, one in each hemisphere, for full sky coverage. The northern site in the Canary Island of La Palma (Spain, 28.76$^{o}$ N, 17.89$^{o}$ W, 2180 m a.s.l.) will host 4 Large Size Telescopes (LSTs) and 15 Medium Size Telescopes (MSTs). The southern site close to Cerro Paranal (Chile, 24.07$^{o}$ S, 70.30$^{o}$ W, 2150 m a.s.l.) will host 4 LSTs, 25 MSTs and 70 Small Size Telescopes (SSTs). The different telescope types will allow to cover a very wide energy range. The LSTs with their 23 m diameter dishes will be dedicated to the low energy range with a threshold of some tens of GeV, hunting for transients and for the most distant objects. In October 2018 the first LST (LST1) has been inaugurated in the northern site and is currently in the commissioning phase \cite{juan}. LST1 is located at about 100 m from MAGIC\footnote{https://magic.mpp.mpg.de/}, a stereoscopic system of two 17 m telescopes, with an energy threshold of 50 GeV with standard trigger (30 GeV with SumTrigger \cite{sum}). Thanks to the locations of MAGIC and LST1 it will be possible to cross-calibrate the two instruments and furthermore to perform combined observations. In this work we evaluated the performance potentially reached by such combined observations.

\section{Monte Carlo simulation and analysis}
\begin{figure}
\centering
\begin{minipage}{.5\textwidth}
  \centering
  \includegraphics[width=.95\linewidth]{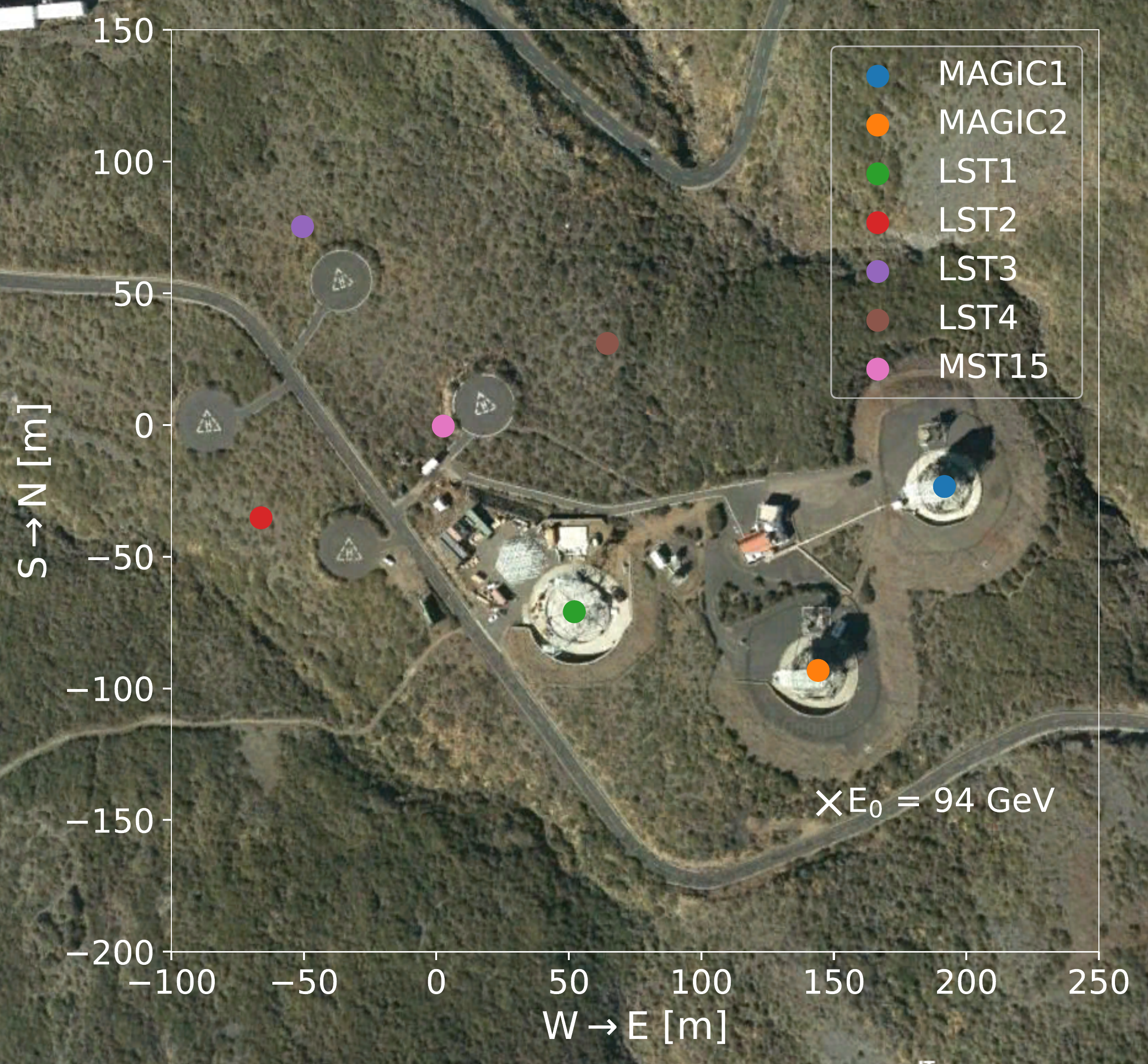}
 % \caption{Simulated Telescope positions (4 LSTs, 2 MAGIC and 1 MST), site of La Palma.}
  % \label{fig:masterarraylayout}
\end{minipage}%
\begin{minipage}{.5\textwidth}
  \centering
  \includegraphics[width=.95\linewidth]{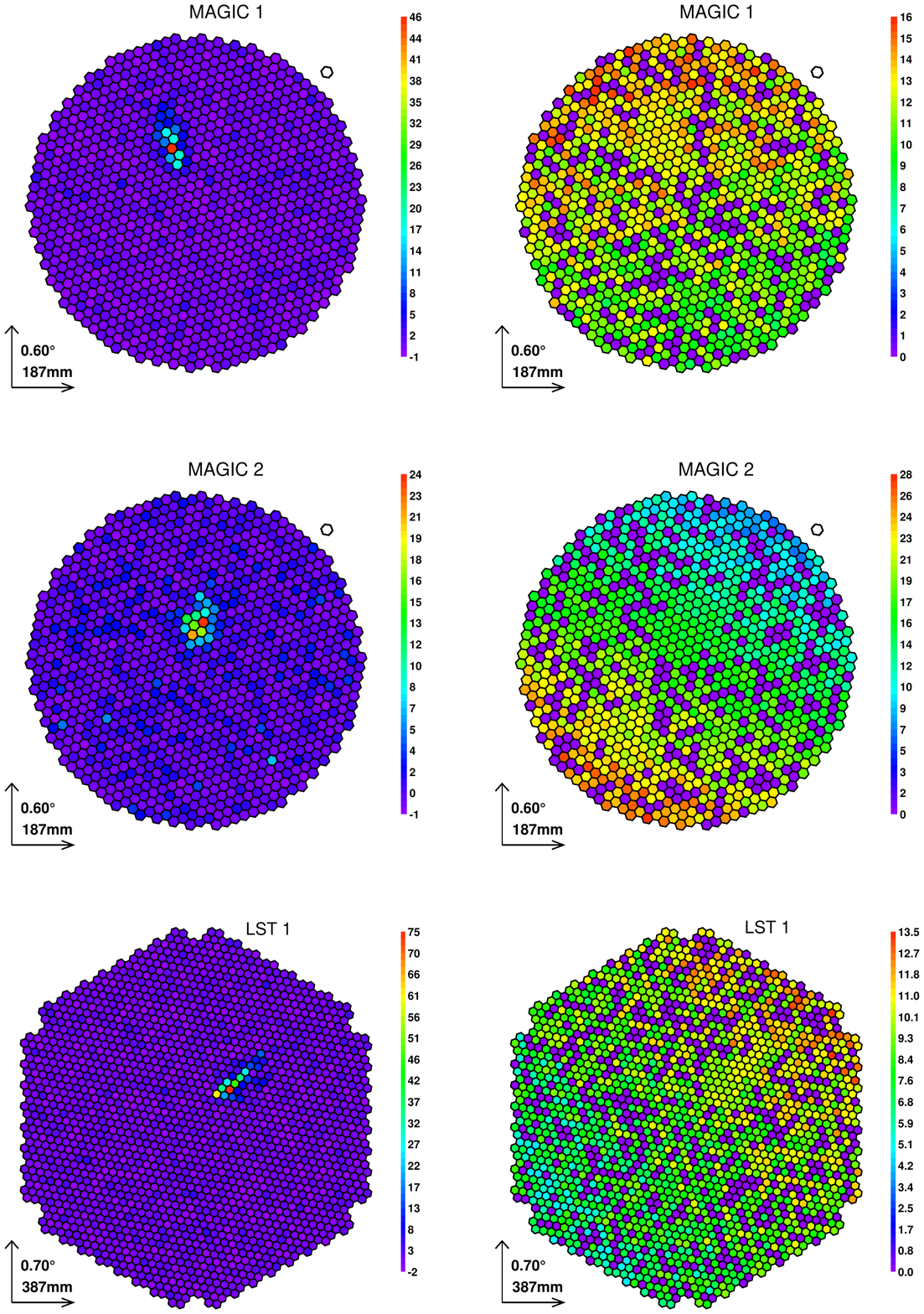}
  %\caption{Images on MAGIC1, MAGIC2 and LST1 cameras of a 94 GeV gamma, with core position shown in fig.\ref{fig:masterarraylayout}}
   %\label{fig:camera-display}
\end{minipage}
\caption{(Left) Simulated telescope positions (4 LSTs, 2 MAGIC and 1 MST), site of La Palma. (Right) Images on MAGIC1, MAGIC2 and LST1 cameras of a 94 GeV gamma, the core position is shown in the left panel with a white cross.}
\label{fig:masterarraylayout}
\end{figure}

The evaluation of the expected performance of an Imaging Atmospheric Cherenkov Telescope (IACT) system is provided by means of large scale Monte Carlo productions. This production was done using the CTA Virtual Organization Grid resources \cite{dirac}.\\
The simulation of the extensive air showers and their Cherenkov light emission was done with \textit{CORSIKA} \cite{corsika}, including specific atmospheric and geomagnetic information for La Palma. For the positions of the simulated telescopes on the ground we have used the actual position of LST1 and of the MAGIC telescopes and we also included all other LSTs of the CTA North array \cite{cumani}, for straightforward comparisons with previous productions, and one MST (fig. \ref{fig:masterarraylayout}). We simulated different primaries, gammas (point-like and diffuse), protons and electrons (diffuse), at zenith angle of 20$^o$ and for 2 pointings (towards North and South), with a spectral index = -2. In table \ref{tab:corsika} the main parameters of the EAS production have been summarized.

\begin{table}[ht]

\centering

\begin{tabular}[t]{lccc}
\hline
&Gamma&Proton&Electron\\
\hline
Energy Range [GeV] & $3 - 330 \cdot 10^3$ & $4 - 600 \cdot 10^3$ & $3 - 330 \cdot 10^3$\\
Radius for core scattering [m] & 700 & 1000 & 1000 \\
Maximum angular distance between & 0 & 6 & 6 \\
telescope axis and shower direction [deg] & & & \\
Number of simulated showers & $1.1\cdot 10^9$ & $2.4\cdot 10^{10}$& $2.0\cdot 10^{10}$\\
\hline
\end{tabular}
\caption{\textit{CORSIKA} production summary. The full production occupies a volume of 280 TB.}
\label{tab:corsika}
\end{table}%

\newpage
%\begin{figure}
% \begin{center}
%  \includegraphics[width=0.6\linewidth]{layout_corsika_prod.png}
%  \caption{Simulated Telescope positions (4 LSTs, 2 MAGIC and 1 MST), site of La Palma.}
%  \label{fig:masterarraylayout}
% \end{center}
%\end{figure}

The simulation of the telescopes was done using \textit{sim$\_$telarray} \cite{simtel}, applied for the first time to simulate also the MAGIC telescopes. The implementation of the MAGIC telescopes has been extensively checked and tuned to reproduce the results of the standard MC software used by the MAGIC Collaboration \cite{pratik}.\\
The simulated events have been reconstructed and analyzed with the MARS \cite{mars} analysis chain, a software developed and used by the MAGIC Collaboration, adapted to handle CTA simulations. For each pixel the charge is extracted by means of a fixed-width sliding window, the image is then cleaned with a standard two-levels tailcut algorithm and the image is parameterized accordingly to classical second moment Hillas analysis \cite{hillas}. Basic quality cuts have been applied, e.g.: image centroid within 0.8 of camera radius and image size larger than 50 photoelectrons. The reconstruction of the event's arrival direction is done with lookup tables, while the energy reconstruction and the background suppression are done by means of Random Forests.
The expected background rates have been calculated using the primary spectra measured by cosmic rays experiments (BESS for protons and Fermi for electrons); the contribution of heavier nuclei after gamma-ray selection cuts is negligible. 
The performance has been calculated for an on-axis point-like source.
From the complete simulated layout (7 telescopes) it is possible to extract any sublayout (e.g.: only MAGIC, 4 LSTs, MAGIC and LST1). An example of an event (E$_0$=94 GeV, core position shown in the map), as seen by the MAGIC1, MAGIC2 and LST1 cameras, is shown in the right panel of fig.\ref{fig:masterarraylayout} (pixels' charge on the left, timing on the right). 

\section{Performance of combined observations}

\begin{figure}
 \begin{center}
  \includegraphics[width=0.8\linewidth]{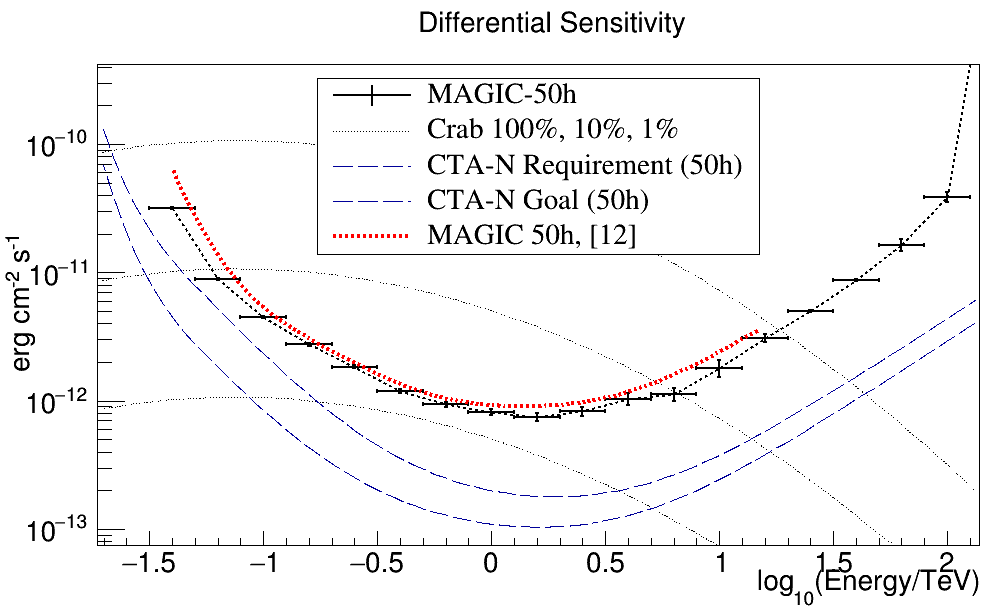}
  \caption{Simulated MAGIC differential sensitivity (50h) and measured one \cite{magic-perf}.}
  \label{fig:magic-lit}
 \end{center}
\end{figure}

The results have been checked for consistency with previous CTA productions (for instance the subarray with 4 LSTs) and, in the case of MAGIC, the obtained sensitivity has been compared to the measured one, see fig. \ref{fig:magic-lit}. The good agreement confirms that the simulation of the MAGIC telescopes and the analysis chain are reliable.\\
 
\begin{figure}
 \begin{center}
  \includegraphics[width=0.8\linewidth]{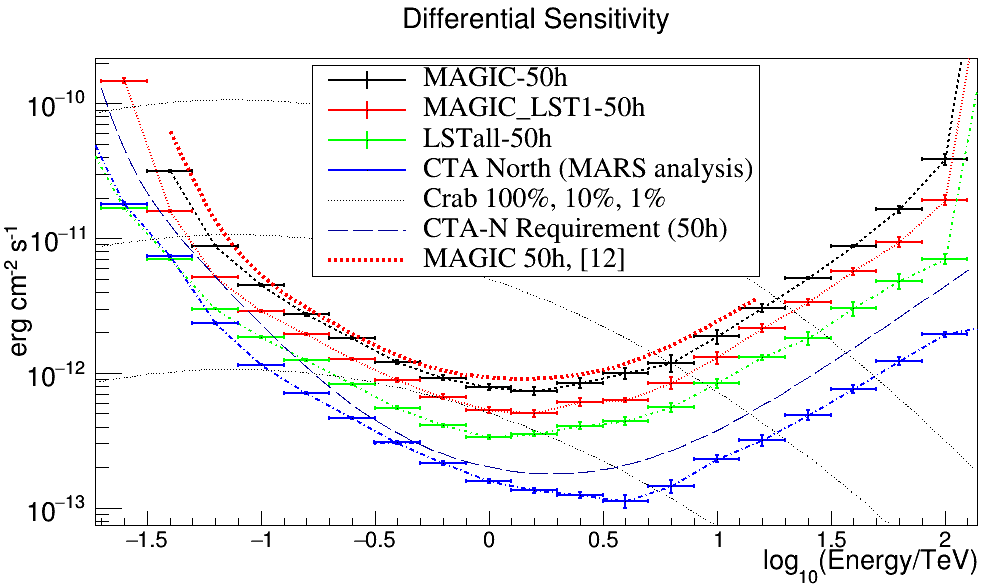}
  \caption{Differential sensitivities for point-like source in 50 hours, for MAGIC, MAGIC-LST1 combined observations and 4 LSTs. As references are also shown the CTA North requirement and simulated sensitivity (MARS analysis result).}
  \label{fig:magic-lst1_sens}
 \end{center}
\end{figure}

\begin{figure}
 \begin{center}
  \includegraphics[width=0.5\linewidth]{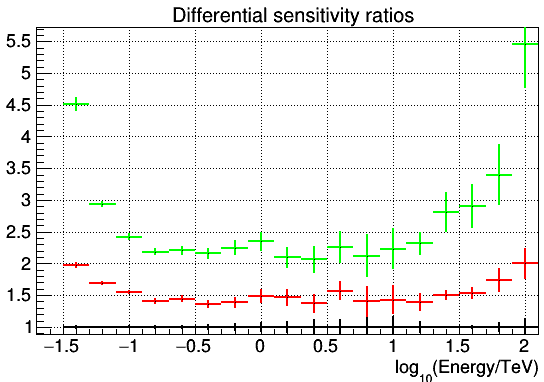}
  \caption{Differential sensitivities ratios (larger is better), for MAGIC-LST1 combined observations (red) and 4 LSTs (green), with respect to MAGIC (black).}
  \label{fig:magic-lst1_ratios}
 \end{center}
\end{figure}

The point source differential sensitivity is shown in fig. \ref{fig:magic-lst1_sens} for MAGIC, MAGIC and LST1 combined observations and for the 4 LSTs (labelled as ''LST all''). The differential sensitivity ratios with respect to MAGIC are shown in fig. \ref{fig:magic-lst1_ratios}.  The simulated trigger condition for MAGIC-LST1 combined observation was the standard MAGIC stereo trigger and mono-trigger for LST1; in the analysis step a request of at least 2 images is set, so it may happen that one image from a MAGIC telescope and the other from LST1 are used, in case the other MAGIC telescope has triggered but its image has been discarded by the analysis quality cuts. 

\begin{figure} \vspace{-1pc}
 \begin{center}
  \includegraphics[width=0.5\linewidth]{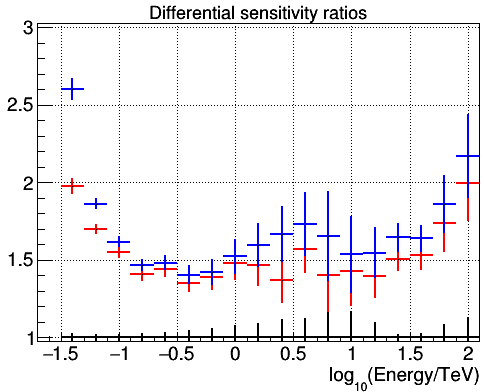}
  \vspace{-1pc}
  \caption{Differential sensitivities ratios for MAGIC-LST1 combined observations with hypothetical hardware trigger (any 2 out of 3, in blue) and the software-only combined MAGIC-LST1 trigger (red), with respect to MAGIC (black).}
  \label{fig:magic-lst-hwtrig}
 \end{center}\vspace{-1pc}
\end{figure}

We explored also the performance of a hypothetical hardware trigger between MAGIC and LST1 (any 2 out of the 3 telescopes). The comparison with a software-only combined trigger (i.e.: events are collected by MAGIC in stereo and LST1 in mono and are put together by means of time tags) is shown in fig. \ref{fig:magic-lst-hwtrig}.  

\section{Conclusions}
In the near future MAGIC and LST1 will perform simultaneous observations with the purpose of supporting LST1 commissioning. Beside cross-calibration purposes it is interesting to evaluate the performance that can be reached by combined analysis of those data. After having checked the reliability of our simulations comparing our results with previous CTA MC productions and MAGIC published sensitivity, we have studied the expected performance of MAGIC and LST1 combined observations. It has been shown that already combining the events by means of simple time tagging and including the images in a common analysis framework will provide on average a factor 1.5 better sensitivity with respect to MAGIC alone observations (even more in some energy bins). We have also studied the performance that could be achieved with a hardware trigger (2/3) including the 2 MAGIC telescopes and LST1, finding a significant improvement of the sensitivity in the first energy bins.

\newpage
\noindent
\textbf{Acknowledgements} We gratefully acknowledge financial support from the agencies and organizations listed here: http://www.cta-observatory.org/consortium$\_$acknowledgments

\end{document}